\documentclass[11pt]{iopart}
\usepackage{graphicx}
\usepackage{epsfig}
\usepackage{multirow}      
\usepackage{xspace}
\usepackage{subfigure}

\usepackage{iopams}
\usepackage{color}

\newcommand{\text}{\rm }

\newcommand{\cmSqSec}{cm$^{-2}$~s$^{-1}$}
\newcommand{\HeFuse}{D$(p,\gamma)^{3}$He}
\newcommand{\LiSeven}{$^{7}$Li$^{*}$$\rightarrow^{7}$Li$+a$}

\newcommand{\CERN}{European Organization for Nuclear Research (CERN),CH-1211 Gen\`eve 23, Switzerland}
\newcommand{\Saclay}{DAPNIA, Centre d'\'Etudes Nucl\'eaires de Saclay, Gif-sur-Yvette, France}
\newcommand{\Darmstadt}{Technische Universit\"at Darmstadt, Institut f\"{u}r Kernphysik, Schlossgartenstrasse 9, 64289 Darmstadt, Germany}
\newcommand{\MPE}{Max-Planck-Institut f\"{u}r extraterrestrische Physik, Giessenbachstrasse, 85748 Garching, Germany}
\newcommand{\Zaragoza}{Instituto de F\'{\i}sica Nuclear y Altas Energ\'{\i}as, Universidad de Zaragoza, Zaragoza, Spain }
\newcommand{\Chicago}{Enrico Fermi Institute and KICP, University of Chicago, Chicago, IL, USA}
\newcommand{\Thessaloniki}{Aristotle University of Thessaloniki, Thessaloniki, Greece}
\newcommand{\Athens}{National Center for Scientific Research ``Demokritos'', Athens, Greece}
\newcommand{\Freiburg}{Albert-Ludwigs-Universit\"{a}t Freiburg, Freiburg, Germany}
\newcommand{\INR}{Institute for Nuclear Research, Russian Academy of Sciences, Moscow, Russia}
\newcommand{\Vancouver}{Department of Physics and Astronomy, University of British Columbia, Vancouver, Canada }
\newcommand{\Frankfurt}{Johann Wolfgang Goethe-Universit\"at, Institut f\"ur Angewandte Physik, Frankfurt am Main, Germany}
\newcommand{\MPI}{Max-Planck-Institut f\"{u}r Physik (Werner-Heisenberg-Institut), F\"ohringer Ring 6, 80805 M\"unchen, Germany}
\newcommand{\Zagreb}{Rudjer Bo\v{s}kovi\'{c} Institute, Bijeni\v{c}ka cesta 54, P.O.Box 180, HR-10002 Zagreb, Croatia}
\newcommand{\Pisa}{Scuola Normale Superiore, Pisa, Italy}

\newcommand{\Alberta}{Department of Physics, University of Alberta, Edmonton, T6G2G7, Canada}
\newcommand{\BNL}{Brookhaven National Laboratory, NY, USA}

\newcommand{\Patras}{University of Patras, Patras, Greece}
\newcommand{\Lyon}{Inst. de Physique Nucl\'eaire, Lyon, France}

\newcommand{\Bochum}{Ruhr-Universit\"{a}t Bochum, Bochum, Germany}
\newcommand{\Karlsruhe}{Institut f\"{u}r Experimentelle Kernphysik, Universit\"{a}t Karlsruhe, Karlsruhe, Germany}
\newcommand{\Fermilab}{Fermi National Accelerator Laboratory, Batavia, IL, USA}
\newcommand{\StanfordSLAC}{Stanford University and SLAC National Accelerator Laboratory, Stanford, CA, USA}

\begin{document}

\title[Search for solar axions from nuclear decays with CAST]{Search for solar axion emission from $^{7}$Li and \HeFuse\ nuclear decays with the CAST $\gamma$-ray calorimeter}

\author{S.~Andriamonje$^{1}$,
S.~Aune$^{1}$,
D.~Autiero$^{2,16}$, 
K.~Barth$^{2}$, 
A.~Belov$^{3}$,
B.~Beltr\'an$^{4,17}$, 
H.~Br\"auninger$^{5}$, 
J.~M.~Carmona$^{4}$, 
S.~Cebri\'an$^{4}$, 
J.~I.~Collar$^{6}$, 
T.~Dafni$^{1,11,18}$,
M.~Davenport$^{2}$,
L.~Di~Lella$^{2,19}$,
C.~Eleftheriadis$^{7}$, 
J.~Englhauser$^{5}$,
G.~Fanourakis$^{8}$, 
E.~Ferrer-Ribas$^{1}$, 
H.~Fischer$^{9}$,
J.~Franz$^{9}$,
P.~Friedrich$^{5}$, 
T.~Geralis$^{8}$, 
I.~Giomataris$^{1}$, 
S.~Gninenko$^{3}$,
H.~G\'omez$^{4}$, 
M.~Hasinoff$^{10}$, 
F.~H.~Heinsius$^{9}$,
D.~H.~H.~Hoffmann$^{11}$,
I.~G.~Irastorza$^{1,4}$, 
J.~Jacoby$^{12}$,
K.~Jakov\v{c}i\'{c}$^{13}$, 
D.~Kang$^{9,21}$, 
K.~K\"onigsmann$^{9}$,
R.~Kotthaus$^{14}$, 
M.~Kr\v{c}mar$^{13}$, 
K.~Kousouris$^{8,22}$,
M.~Kuster$^{11,5}$, 
B.~Laki\'{c}$^{13}$, 
C.~Lasseur$^{2}$,
A.~Liolios$^{7}$, 
A.~Ljubi\v{c}i\'{c}$^{13}$,
G.~Lutz$^{14}$, 
G.~Luz\'on$^{4}$, 
D.~W.~Miller$^{6,23}$,
J.~Morales$^{4}$,  
A.~Ortiz$^{4}$,
T.~Papaevangelou$^{1,2}$, 
A.~Placci$^{2}$,
G.~Raffelt$^{14}$, 
H.~Riege$^{11}$,
A.~Rodr\'iguez$^{4}$, 
J.~Ruz$^{4,24}$, 
I.~Savvidis$^{7}$,
Y.~Semertzidis$^{15,25}$,
P.~Serpico$^{2,14}$, 
L.~Stewart$^{2}$, 
J.~D.~Vieira$^{6}$,
J.~Villar$^{4}$, 
J.~Vogel$^{9}$,
L.~Walckiers$^{2}$ and 
K.~Zioutas$^{2,15}$
(CAST Collaboration)}

\address{$^1$ \Saclay}
\address{$^2$ \CERN}
\address{$^3$ \INR}
\address{$^4$ \Zaragoza}
\address{$^5$ \MPE}
\address{$^6$ \Chicago}
\address{$^7$ \Thessaloniki}
\address{$^8$ \Athens}
\address{$^{9}$ \Freiburg}
\address{$^{10}$ \Vancouver}
\address{$^{11}$ \Darmstadt}
\address{$^{12}$ \Frankfurt}
\address{$^{13}$ \Zagreb}
\address{$^{14}$ \MPI}
\address{$^{15}$ \Patras}

\vspace{1cm}

\address{$^{16}$ Present address: \Lyon}
\address{$^{17}$ Present address: \Alberta}
\address{$^{18}$ Present address: \Zaragoza}
\address{$^{19}$ Present address: \Pisa}
\address{$^{20}$ Present address: \Bochum}
\address{$^{21}$ Present address: \Karlsruhe}
\address{$^{22}$ Present address: \Fermilab}
\address{$^{23}$ Present address: \StanfordSLAC}
\address{$^{24}$ Present address: \CERN}
\address{$^{25}$ Present address: \BNL}

\ead{David.Miller@slac.stanford.edu}

\begin{abstract}

We present the results of a search for a high-energy axion emission signal from $^{7}$Li (0.478~MeV) and \HeFuse\ (5.5~MeV) nuclear transitions using a low-background $\gamma$-ray calorimeter during Phase I of the CAST experiment. These so-called ``hadronic axions'' could provide a solution to the long-standing \textit{strong-CP} problem and can be emitted from the solar core from nuclear M1 transitions. This is the first such search for high-energy pseudoscalar bosons with couplings to nucleons conducted using a helioscope approach. No excess signal above background was found.

\vspace{3mm}                              
\begin{flushleft} \textbf{Keywords}: axions, axion-photon coupling, axion-nucleon coupling, hadronic axions
\end{flushleft}                             

\end{abstract}


\maketitle

\newpage

\section{Introduction}

The observed $CP$ invariance in the strong interactions is not \textit{a priori} expected, as 't Hooft pointed out~\cite{tHoof76}, and has been named the \emph{strong-$CP$ problem}. Nonperturbative effects in the theory give rise to a $CP$ violating ``$\theta$'' term which appears in the QCD lagrangian as
\begin{equation}\label{QCDLag}
    \mathcal{L}_{\theta}=\theta\frac{\alpha_S}{8\pi}G_{\mu\nu}\tilde G^{\mu\nu}.
\end{equation}
Here, $G_{\mu\nu}$ is the gluon field strength and $\tilde G^{\mu\nu}=\frac{1}{2}\epsilon_{\mu\nu\rho\sigma}G^{\rho\sigma}$ its dual. The apparent $CP$ invariance of QCD derives from the fact that $\theta$ is measured to be vanishingly small via the neutron electric dipole moment, for which the current upper limit is $|d_n|<6.3\times10^{-26}$ $e$ cm~\cite{harr99}. This limit on $d_n$ implies an upper limit $\theta<10^{-10}$~\cite{crew79}.

In 1977, Peccei and Quinn proposed a physical origin for $\theta=0$ by introducing a global $U(1)$ chiral symmetry~\cite{pec77}, often referred to as $U(1)_{\rm PQ}$. The parameter $\theta$ thus becomes a dynamical variable that is forced to zero when the potential is minimized. Weinberg and Wilczek showed that such a solution implies the existence of a new particle, the axion, and that such a particle can have couplings to quarks, nucleons, leptons and photons~\cite{wein78,wil78}.

The axion was first thought to have couplings on the order of the weak scale~\cite{wein78} and a mass of $\sim$200 keV. Experimental evidences against this coupling strength and mass range, most notably through limits on the magnetic moment of the muon, kaon decay and quarkonium studies, prompted the idea of ``invisible'' axions.
There are two classes of invisible axion models: KSVZ (Kim, Shifman, Vainshtein, and Zakharov)~\cite{kim79,shifman80} and DFSZ (Dine, Fischler, Srednicki, and Zhitnitski\u{\i})~\cite{dfs81,zhi80}. In the former axion models, commonly referred to as hadronic axion models, couplings to leptons are strongly suppressed.
Since couplings to nucleons and photons remain, detection of the hadronic axion is still possible. Other models have also been proposed with suppressed axion-photon coupling~\cite{kaplan85}, but we focus here on axion models which include couplings to both photons and nucleons.

Most of the experiments searching for axions or similar pseudoscalar bosons~\cite{raf88} have been relying on the coupling to two photons i.e. Primakoff effect~\cite{prim51}. The axion-photon coupling is given by the effective Lagrangian
\begin{eqnarray}  \label{eq:axion_photon_lagrangian}
{\cal{L}}_{\rm{a\gamma}}=-\frac{1}{4}\, g_{\rm{a\gamma}}\,F^{\mu\nu}\,\tilde{F}_{\mu\nu}\,a=g_{\rm{a\gamma}}\, {\bf{E}}\cdot{\bf{B}}\,a\,,
\end{eqnarray}
where $a$ is the axion field, $F^{\mu\nu}$ the electromagnetic field strength tensor, and $\tilde{F}_{\mu\nu}$ its dual, $\bf{E}$ the electric and $\bf{B}$ the magnetic field of the coupling photons. The effective axion-photon coupling constant $g_{\rm{a\gamma}}$ is given by
\begin{eqnarray} \label{eq:gagamma}
g_{\rm{a\gamma}}=\frac{\alpha}{2\pi f_{\rm{a}}} \left[ \frac{E}{N}-\frac{2\,(4+z+w)}{3\,(1+z+w)}\right] =
\frac{\alpha}{2\pi f_{\rm{a}}} \left( \frac{E}{N}-1.95\pm 0.08\right) \, ,
\end{eqnarray}
where $f_{\rm a}$ is the Peccei-Quinn symmetry breaking scale, $z=m_{\rm u}/m_{\rm d}=0.56$ and $w=m_{\rm u}/m_{\rm s}=0.028$ are the quark mass ratios. Here $E$ and $N$ are the model dependent coefficients of the electromagnetic and color anomaly of the axial current associated with the $U(1)_{\rm PQ}$ symmetry, respectively. Frequently cited axion models use $E/N=0$~\cite{kim79, shifman80} or $E/N=8/3$~\cite{dfs81,zhi80} but, in general, $E/N$ can take different
values depending on the specific model details.

In the Primakoff process an axion couples to a virtual photon in an electromagnetic field and converts to a real photon, or vice-versa. The conversion probability in vacuum, in a magnetic field of length $L$ and strength $B$, depends on the axion-photon coupling constant $g_{{\rm a}\gamma}$ and the momentum transfer between the axion and the photon $q$ as given in~\cite{vanbib89}
\begin{equation}\label{ConvProb}
    P_{\mathrm{a}\rightarrow\gamma}(L)=
    \left( \frac{g_{\mathrm{a}\gamma}BL}{2} \right)^{2}
\frac{4}{q^{2} L^{2}} \sin^{2} \left( \frac{qL}{2} \right).
\end{equation}
In addition, this probability depends implicitly on the axion mass $m_{\rm a}$ and the axion energy $E_{\rm a}$ through the relation $q=m_{\rm a}^2/2E_{\rm a}$.
For $qL\lesssim1$ the conversion is coherent and equation~(\ref{ConvProb}) reduces to $(g_{\mathrm{a}\gamma}BL/2)^2$.

\begin{figure}[!h]
\begin{center}
\includegraphics[width=0.60\textwidth]{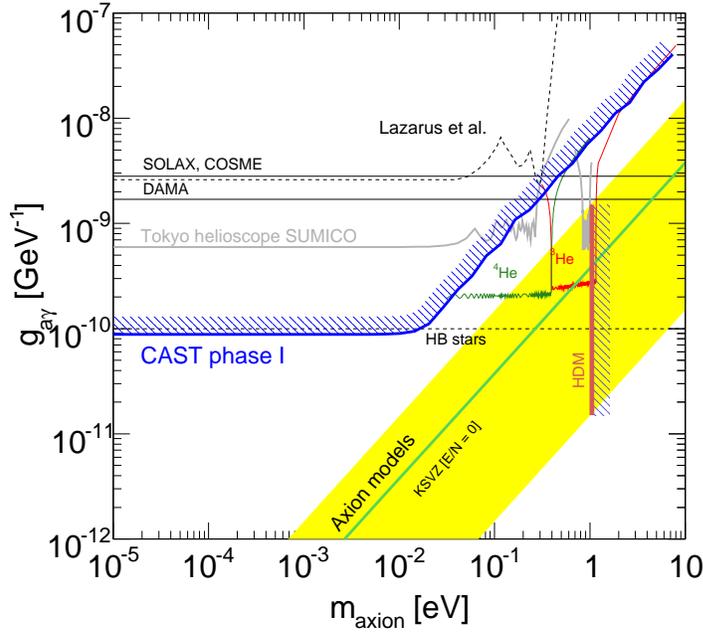}
\caption{\label{castlimits} Exclusion plots in the axion-photon coupling versus the axion mass plane. The limit achieved on Primakoff axions from the X-ray detectors by the previous Phase~I of the CAST experiment ~\cite{Andriamonje:2007ew} (updated limits consistent with expectations are reported in~\cite{arik:2009}) is compared with other constraints (Lazarus et al.~\cite{laz92}, SOLAX~\cite{avignone98}, COSME~\cite{morales02}, DAMA~\cite{Ber01}, Tokyo helioscope experiment SUMICO~\cite{mori98} and HB stars~\cite{raf96}). The vertical red line (HDM) is the hot dark matter limit for hadronic axions $m_{\rm a}<1.05~{\rm eV}$~\cite{Han05} inferred from observations of the cosmological large-scale structure. The yellow band represents typical theoretical models with $\left|E/N-1.95\right|$ in the range 0.07--7 while the green solid line corresponds to the case when $E/N=0$ is assumed, as in the KSVZ model~\cite{kim79,shifman80}.}
\end{center}
\end{figure}

Axion-photon mixing permits a variety of production mechanisms and detection techniques, many of which were first pointed out by Sikivie in 1983~\cite{sik83}. Magnetically induced vacuum birefringence~\cite{zavattini86}, stellar and terrestrial magnetic fields, pulsar magnetic fields, and resonant cavities~\cite{sik83} all provide methods for the production or detection of axions. Experimental and astrophysical limits on the axion are typically stated in terms of the photon coupling, $g_{\mathrm{a}\gamma}$, versus mass, $m_{\mathrm{a}}$, and recent experimental and cosmological limits are shown in figure~\ref{castlimits}.

Due to the axion coupling to nucleons, there are additional components of solar axions
emitted in nuclear de-excitations and reactions. The energy of these mono-energetic axions
corresponds to the energy of the particular process. In this paper we present the results
of our search for mono-energetic solar axions which may be emitted from $^{7}$Li$^*$ de-excitation and
\HeFuse\ reaction by using the CERN Axion Solar Telescope (CAST) setup. To detect photons
coming from conversion of these axions in the CAST magnet, we used high-energy photon
calorimeter that was mounted on one end of the magnet during the CAST phase I.

In section~\ref{Nuclear} nuclear axion couplings are discussed and the signal expected from nuclear axion emission in the Sun from the first excited level of $^{7}$Li (0.478~MeV) and reaction \HeFuse\ (5.5~MeV) is described. In sections~\ref{datasection} and~\ref{Analysis} the data selection, systematics and analysis of the data are then presented in detail.

\section{Nuclear axion emission in hadronic axion models\label{Nuclear}}

\subsection{Axion-nucleon coupling}

Coupling to nucleons occurs through the spin operator $\mathbf{\sigma}$~\cite{avignone88} and because axions carry spin-parity $J^P=0^-,1^+,2^-,...$ nuclear deexcitation via axion emission occurs predominantly via M1 magnetic nuclear transitions. Several channels exist for solar axion emission via these transitions~\cite{raf82,Mor95,Krc98,krc01,Jak04,Andriamonje:2009M1}.
Here we focused our attention on the thermonuclear fusion reaction ${\rm p}+{\rm d}$$\rightarrow$$^{3}{\rm He}+a \,{\rm(5.5~MeV)}$ and associated reaction chain  $^{3}{\rm He}(\alpha,\gamma)^{7}{\rm Be}({\rm e}^{-},\nu_{\rm e})^{7}{\rm Li}^{*}$$\rightarrow$$^{7}{\rm Li}+a\,{\rm(0.478~MeV)}$ as a source of solar mono-energetic axions.


The branching ratio for axion emission via M1 transitions is directly calculated as~\cite{avignone88}
\begin{eqnarray} \label{brratio}
    \frac{\Gamma_{\rm a}}{\Gamma_{\gamma}} &=& \frac{1}{2\pi\alpha}\left(\frac{k_{\rm a}}{k_{\gamma}}\right)^3\left(\frac{1}{1+\delta^2}\right) \left[\frac{g_{0}\beta+g_{3}}{(\mu_{0}-1/2)\beta+\mu_{3}-\eta}\right]^2.
\end{eqnarray}
Here, $k_{\rm a}$ and $k_{\gamma}$ are the axion and photon momenta, respectively, and will both be approximately equal to the decay channel energy. In addition, $\alpha=1/137$, $\delta$ is the $E2/M1$ mixing ratio, $\mu_0\approx0.88$ and $\mu_3\approx4.71$ are the isoscalar and isovector nuclear magnetic moments, respectively, in nuclear magnetons.

The remaining terms $\eta$ and $\beta$ are nuclear structure dependent parameters which depend directly on the initial and final state nuclear wavefunctions. It is via these terms that the \textit{decay-channel specific} axion-physics is expressed, and are given by~\cite{avignone88}

\begin{displaymath}
    \eta=-\frac{\langle \mathbf{J}_{f}|| \sum_{i=1}^{\rm A} \mathbf{l}(i)\tau_{3}(i) ||\mathbf{J}_{i}\rangle}{\langle \mathbf{J}_{f}|| \sum_{i=1}^{\rm A} \mathbf{\sigma}(i)\tau_{3}(i) ||\mathbf{J}_{i}\rangle}, \hspace{1cm} \beta=\frac{\langle \mathbf{J}_{f}|| \sum_{i=1}^{\rm A} \mathbf{\sigma}(i) ||\mathbf{J}_{i}\rangle}{\langle \mathbf{J}_{f}|| \sum_{i=1}^{\rm A} \mathbf{\sigma}(i)\tau_{3}(i) ||\mathbf{J}_{i}\rangle},
\end{displaymath}
where $\mathbf{J}_{i}$ and $\mathbf{J}_{f}$ are the angular momenta of 
initial and final state respectively, $\mathbf{l}(i)$ is the nucleon orbital angular
momentum operator, $\mathbf{\sigma}(i)$ represents the spin operator, while $\tau_{3}(i)$
is the isospin operator.

In equation~(\ref{brratio}) we have written the axion-nucleon coupling as $(g_0\beta+g_3)$, where $g_0$ is the isoscalar coupling and $g_3$ the isovector coupling. In the hadronic axion models these are written as~\cite{kaplan85}
\begin{eqnarray} \label{g0}
    g_0 &=& -\frac{m_{\rm N}}{f_{\rm a}}\frac{1}{6}\left[2S+(3F-D)\frac{1+z-2w}{1+z+w} \right] \nonumber\\
    g_3 &=& -\frac{m_{\rm N}}{f_{\rm a}}\frac{1}{2}\left[(D+F)\frac{1-z}{1+z+w}\right], \nonumber\\ \label{g3}
\end{eqnarray}
where $S=0.4$ characterizes the flavour singlet coupling, $F=0.460$ and $D=0.806$ are matrix elements for the SU(3) octet axial vector currents, $m_{\rm N}=0.939$ GeV is the nucleon mass, yielding $g_{0}=-0.21\,{\rm GeV}/f_{\rm a}$ and $g_{3} =-0.17\,{\rm GeV}/f_{\rm a}$.

\subsection{Expected axion flux from \LiSeven}

The decay of the first excited state of $^{7}$Li
\begin{equation} \label{lithiumToPhoton}
    ^{7}\text{Li}^{*}\rightarrow ^{7}\text{Li}+\gamma \hspace{3mm} (0.478 \text {~MeV})
\end{equation}
follows from $^{7}$Be electron capture ($^{7}$Be$+{\rm e}^{-}\rightarrow$ $^{7}\text{Li}^{*}+\nu_{\rm e}$). Because this process can emit an axion of the same energy instead of a $\gamma$-ray and occurs for each neutrino emission, we can use the measured $^{7}$Be neutrino flux, $\Phi_{\nu}^{\rm Be}$, to estimate the differential flux of axions arriving at Earth, as~\cite{krc01}
\begin{eqnarray}\label{flux}
    \frac{d\Phi_{\rm a}}{dE_{\rm a}} &=& \int^{R_{\odot}}_0 d\Phi_{\nu}^{\rm Be}(r)\kappa\frac{\Gamma_{\rm a}}{\Gamma_{\gamma}}\frac{1}{\sqrt{2\pi}\sigma(T)}{\rm exp}\left[-\frac{(E_{\rm a}-E_{\gamma})^2}{2\sigma(T)^2}\right],
\end{eqnarray}
where $R_{\odot}$ is the solar radius, $d\Phi_{\nu}^{\rm Be}(r)$ is the $^{7}$Be neutrino flux at Earth emitted from a solar shell at radius $r$, $\kappa=0.104$ is the branching ratio of the $^{7}$Be electron capture to the first excited state of $^{7}$Li~\cite{toi96}, $\sigma(T)=E_{\gamma}\sqrt{kT/m}$ is a thermal Doppler broadening of the emission line (about 0.2~keV at the solar core), and $\Gamma_{\rm a}/\Gamma_{\gamma}$ is the axion-photon branching ratio. 

We use the values of $\eta$ and $\beta$ calculated in \cite{krc01} as $\eta=0.5$ and $\beta=1.0$, thereby obtaining
\begin{equation}
    \frac{\Gamma_{\rm a}}{\Gamma_{\gamma}}=1.035(g_0+g_3)^{2}.
\end{equation}
Since the resolution of the calorimeter in the energy region around 450 keV is $\sigma_{\rm det}\approx100$ keV, we integrate over the Doppler broadening term and use the total neutrino flux at Earth $\Phi_{\nu}^{\rm Be}=4.86\times10^9$ \cmSqSec~\cite{bachall04}. Thus, we wash out the Doppler term and obtain the total flux of 0.478~MeV solar axions
\begin{eqnarray}
    \Phi_{\rm a}&=&5.23\times10^8(g_0+g_3)^2 \hspace{3mm} {\rm cm}^{-2}{\rm s}^{-1}.
\end{eqnarray}

It is interesting to note the relative insensitivity of $\Phi_{\rm a}$ to the choice of $\eta$ and $\beta$ above. By varying $0.1\le\eta\le0.9$ and $0.6\le\beta\le1.4$, $\Phi_{\rm a}$ varies between $3.2\times10^{8} \leq \Phi_{\rm a}/(g_0\beta+g_3)^2 \leq 8.8\times10^{8}$ (in this range of $\eta$ and $\beta$). Thus, the more critical parameter is the emission rate, which follows the $^7$Be neutrinos, $\Phi_{\nu}^{Be}$. 

\subsection{Expected axion flux from \HeFuse}

Also of interest for hadronic axions is the radiative capture of protons on deuterium, referred to as proton-deuteron fusion~\cite{raf82}. The reaction 

\begin{equation} \label{fusion}
    {\rm p}+{\rm d}\rightarrow^3\text{He}+\gamma \hspace{3mm} (5.5 \text{~MeV})
\end{equation}
occurs at a rate $\phi_{\gamma}$ = 1.7$\times10^{38}$ s$^{-1}$ with only 1/3 of those being M1 transitions. In order to obtain the axion flux expected from this nuclear reaction we must evaluate the axion-photon branching ratio~(\ref{brratio}) using the correct values of parameters $\eta$ and $\beta$, which is made difficult by the fact that (\ref{fusion}) is a 3-body nuclear decay.
However, as pointed out by~\cite{raf82}, (\ref{fusion}) is predominantly isovector, implying that $g_{0}\beta$ is very small and can be neglected. Under this assumption, equation~(\ref{brratio}) becomes~\cite{donnelly78}

\begin{equation} \label{brratio2}
    \frac{\Gamma_{\rm a}}{\Gamma_{\gamma}}=\frac{1}{2\pi\alpha}\left(\frac{k_{\rm a}}{k_{\gamma}}\right)^3\frac{1}{1+\delta^2}\left(\frac{g_{3}}{\mu_{3}}\right)^2.
\end{equation}
Using $\delta$=0, $k_{\rm a}\approx k_{\gamma}=5.5$~MeV, and $\mu_3=4.71$ we have 

\begin{equation} \label{brratio3}
    \frac{\Gamma_{\rm a}}{\Gamma_{\gamma}}=0.98 g_3^2.
\end{equation}
Combining equation~(\ref{brratio3}) and the rate of proton-deuteron fusion reactions characterized as the M1 nuclear transition, we obtain the total flux of 5.5~MeV solar axions
\begin{equation}
    \Phi_{\rm a} = \frac{1}{4\pi d^2_{\odot}} \frac{\phi_{\gamma}}{3} \frac{\Gamma_{\rm a}}{\Gamma_{\gamma}} = 2.03\times10^{10}g_3^2 \hspace{3mm}{\rm cm}^{-2}{\rm s}^{-1},
\end{equation}
where $d_{\odot}$ is the Earth to Sun distance. 

Using the data obtained from the 6 month run in 2004, we can thus evaluate the sensitivity of the CAST $\gamma$-ray calorimeter to these axion signals.

\section{Data \label{datasection}}

\subsection{CAST and the high energy $\gamma$-ray calorimeter \label{sec:CM}}

The CERN Axion Solar Telescope (CAST) utilizes a helioscope design which exploits the increased axion-to-photon conversion probability for increased magnetic field strength and length as given by equation~(\ref{ConvProb}). The refurbished LHC dipole prototype magnet~\cite{zioutas99} produces a nominal magnetic field of $B=9.0$ T over a length of $L=9.26$ m in each of the dipole's two 14.5 cm$^2$ area magnet bores. The full system is mounted on a rotating platform with a vertical range of $\pm8^{\circ}$ and an azimuthal range of $\pm40^{\circ}$. This range of motion allows for 1.5 hours of solar alignment during both sunrise and sunset year-round. The tracking system monitors the alignment of the magnet with the Sun, resulting in a pointing accuracy better than 0.01$^{\circ}$. All remaining time is devoted to background measurements for the low-background X-ray detectors which are installed on both ends of the magnet. Until 2007, a conventional Time Projection Chamber (TPC) covered both magnet bores at one end to detect photons originating from axions during the tracking of the Sun at sunset. It was then replaced by two MICROMEGAS detectors, each attached to one bore. On the other side of the magnet, there is another MICROMEGAS detector covering one bore, and an X-ray mirror telescope with a pn-CCD chip as the focal plane detector at the other bore, both intended to detect photons produced from axions during the sunrise solar tracking. More details about the CAST experiment and detectors can be found in~\cite{Andriamonje:2007ew,arik:2009,Andriamonje:2009M1,zioutas99,Zio05,autiero:07a,kuster:07c,abbon:07a}.

To cover a wide range of potential axion masses, the operation of the CAST experiment is divided into two phases. During the Phase I (2003--2004)~\cite{Andriamonje:2007ew,Zio05} the experiment operated with vacuum inside the magnet bores and the sensitivity was essentially limited to $m_{\rm{a}}<0.02$ eV due to the coherence condition. In the second phase (so-called Phase II) which started in 2005, the magnet bores are filled with a buffer gas in order to extend the sensitivity to higher axion masses. In the first part of this phase (2005--2006) $^{4}$He was used as a buffer gas. By increasing the gas pressure in appropriate steps, axion masses up to $\sim$0.4 eV were scanned and the results of these measurements supersede all previous experimental limits on the axion-photon coupling constant in this mass range~\cite{arik:2009}. To explore axion masses above 0.4\nolinebreak[4] eV, $^{3}$He has to be used because it has a higher vapor pressure than $^{4}$He. This allows for a further increase in gas pressure in the magnet bores to reach axion masses up to about 1 eV in the ongoing second part of Phase II that started in 2007 and is planned to finish by the end of 2010. 

\begin{figure*}
\subfigure[]{\includegraphics[width=0.45\textwidth]{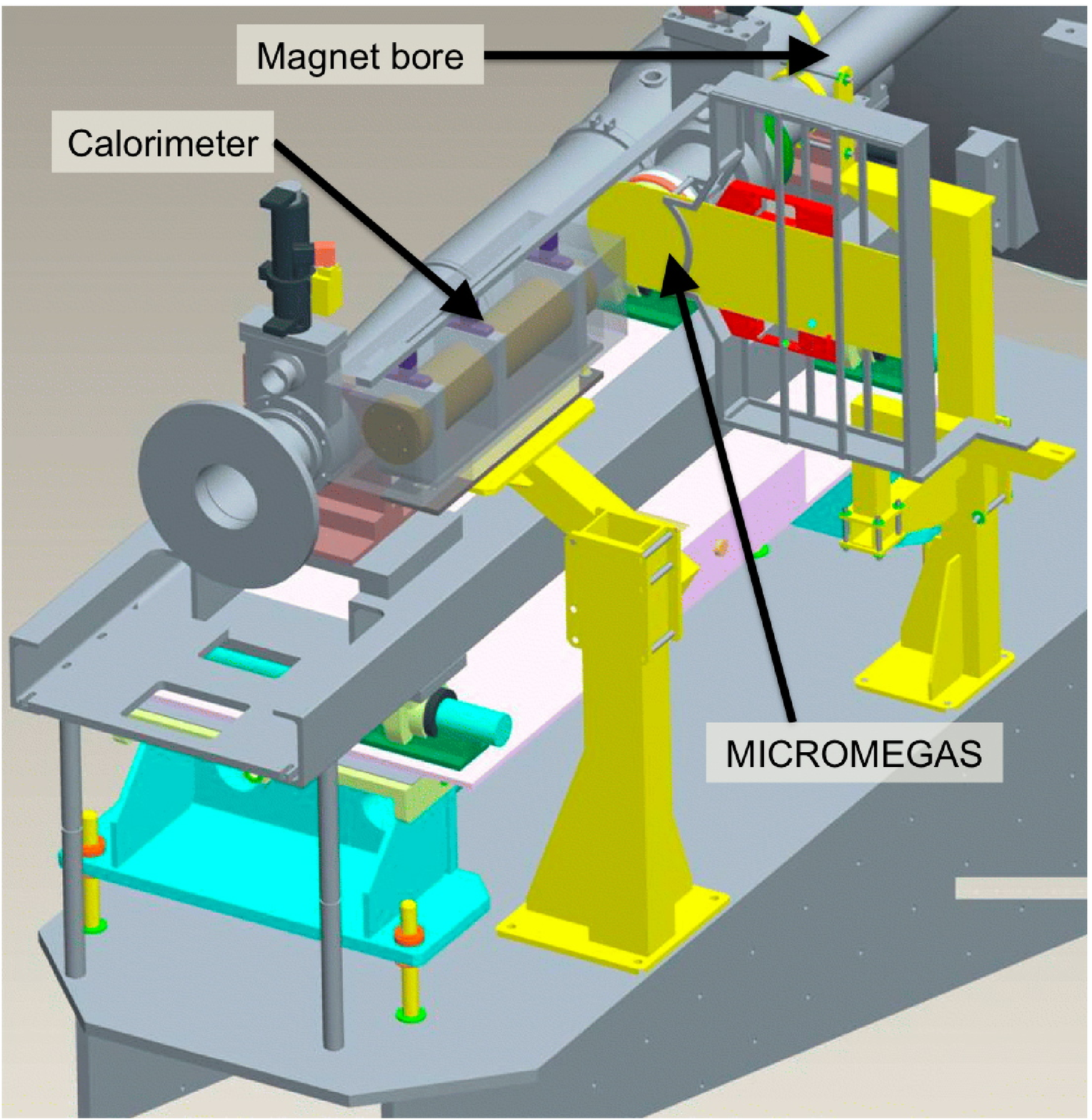}\label{fig:platform}}
\subfigure[]{\includegraphics[width=0.55\textwidth]{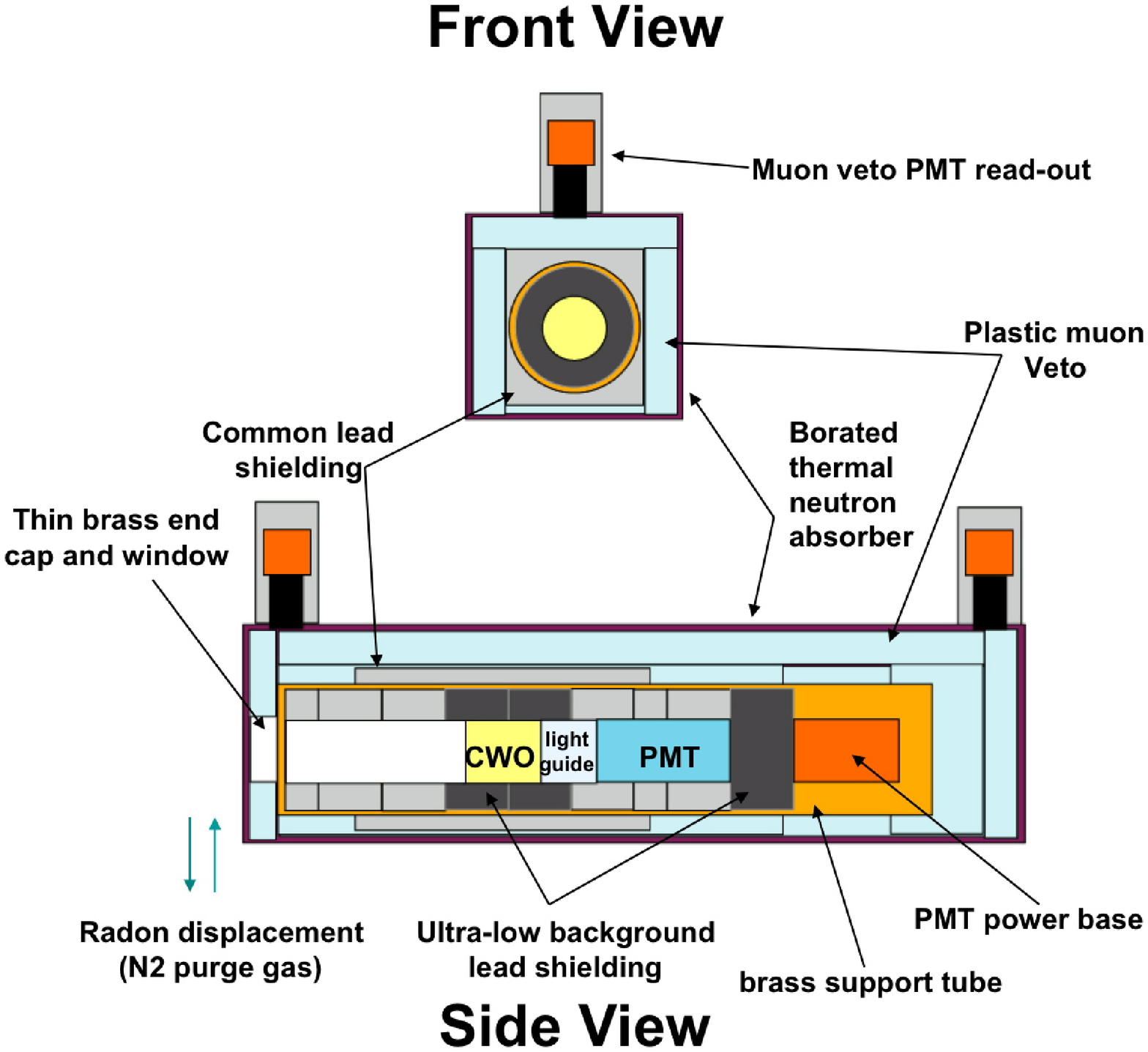}\label{fig:caloSideView}}
\caption{\label{SchematicCalo} \subref{fig:platform} Schematic of the CAST magnet platform depicting the calorimeter placement behind the MICROMEGAS detector and the space constraints limiting the allowed shielding. \subref{fig:caloSideView} Schematic of the CAST high energy $\gamma$-ray calorimeter. The ``tunnel'' design with CWO crystal and PMT surrounded by modest amount of ancient and common lead passive shield. An active muon veto with dual PMT read-out, borated neutron absorber and continuous $N_{2}$ radon purging complement the passive shielding.}
\end{figure*}


Axions emitted in a particular nuclear decay channel will be essentially mono-energetic compared to the Primakoff spectrum expected from plasma processes. However, several candidate processes exist and so the corresponding mono-energetic axion lines are expected to be in the range from tens of keV to many~MeV. The expected axion signal from any single channel is thus a collimated beam of similarly mono-energetic $\gamma$-rays (a ``peak'') from the magnet bore during periods of solar alignment.

The calorimeter consists of a cadmium tungstate scintillating crystal (CdWO$_4$ or CWO), which is also the type used in neutrinoless double-beta decay searches~\cite{faz97}. CWO offers good stopping power for $\gamma$-ray photons, very low internal radioactivity, good energy resolution and excellent pulse shape discrimination characteristics (see section~\ref{cuts}). The crystal is optically coupled to a light guide and photomultiplier tube (PMT) which is placed inside a lead-shielded cylindrical brass tube. This ``tunnel'' design maximizes signal acceptance and background rejection while respecting the space and weight limitations on the CAST detector platform (see figure~\ref{SchematicCalo}(b)). These constraints also limit the allowed thickness (2.5 cm) of ancient and common lead shielding, which results in an elevated environmental background component compared to that achievable with fewer constraints. An active scintillating plastic muon veto, environmental radon purging with constant N$_2$ flow, a borated thermal neutron absorber, and a low-background PMT complement the minimalist passive shielding design. 

The large dynamic range and high stopping power for photons are necessary to achieve a good efficiency at high energies for a generic search. Detector components, shielding materials, and data processing were all designed in order to reduce the environmental backgrounds. Pulse shape discrimination (PSD) further reduces noise and events due to internal radioactive contaminations in the crystal. Finally, an LED pulser provides livetime monitoring. These square pulses are recorded and subsequently removed prior to analysis.

Although similar searches have been conducted in beam dump experiments~\cite{bjorken88,zehn82}, accelerators and terrestrial nuclear processes~\cite{minowa93}, CAST is the first high-energy-axion search using a helioscope. The order of magnitude increase in axion-to-photon conversion probability over previous helioscope searches and the increases intensity of axion emission from the Sun as compared to accelerator searches makes the CAST calorimeter a very sensitive probe of low-mass pseudoscalars. Because axions serve as merely one example of such particles, a high-energy search should not be limited to only axions but should consider anomalous excess during solar tracking events generally~\cite{raf88,raf96}.

The data and results presented in this paper were obtained from CAST Phase I.
\subsection{Pulse shape discrimination and particle identification\label{cuts}}

Following a method similar to~\cite{faz97}, we have developed and applied a pulse shape discrimination (PSD) algorithm which exploits the distinct pulse shape characteristics of the CWO crystal in response to incident particle type. This algorithm relies on the difference in response for nuclear recoils and minimum ionizing particles due to the different mechanisms of scintillation for these two types of excitations. The long $\sim$20$\mu$s decay of CWO results from the presence of 2-4 separate decay constants~\cite{faz97} and enhances these differences and allows for efficient PSD. 

Particle calibrations were performed using alpha's ($\alpha$'s), neutrons ($n$'s), and photons ($\gamma$'s) to create a weighting factor using a statistical average over many pulse shapes for each calibration source. The radioactive sources used for the calibrations represent a spectrum of energies: $^{241}$Am ($\alpha$'s), Am/Be and $^{252}$Cf ($n$'s and $\gamma$'s), and multiple $\gamma$ sources (0.511, 0.662, 0.835, 1.173, 1.333, 1.836 MeV). Particle ``templates'' are formed with these data and are used to create a spectral weighting factor. The PSD algorithm is then applied to the data to remove backgrounds and noise. 

\begin{figure*}
\begin{center}
\subfigure[]{\includegraphics[width=0.45\textwidth]{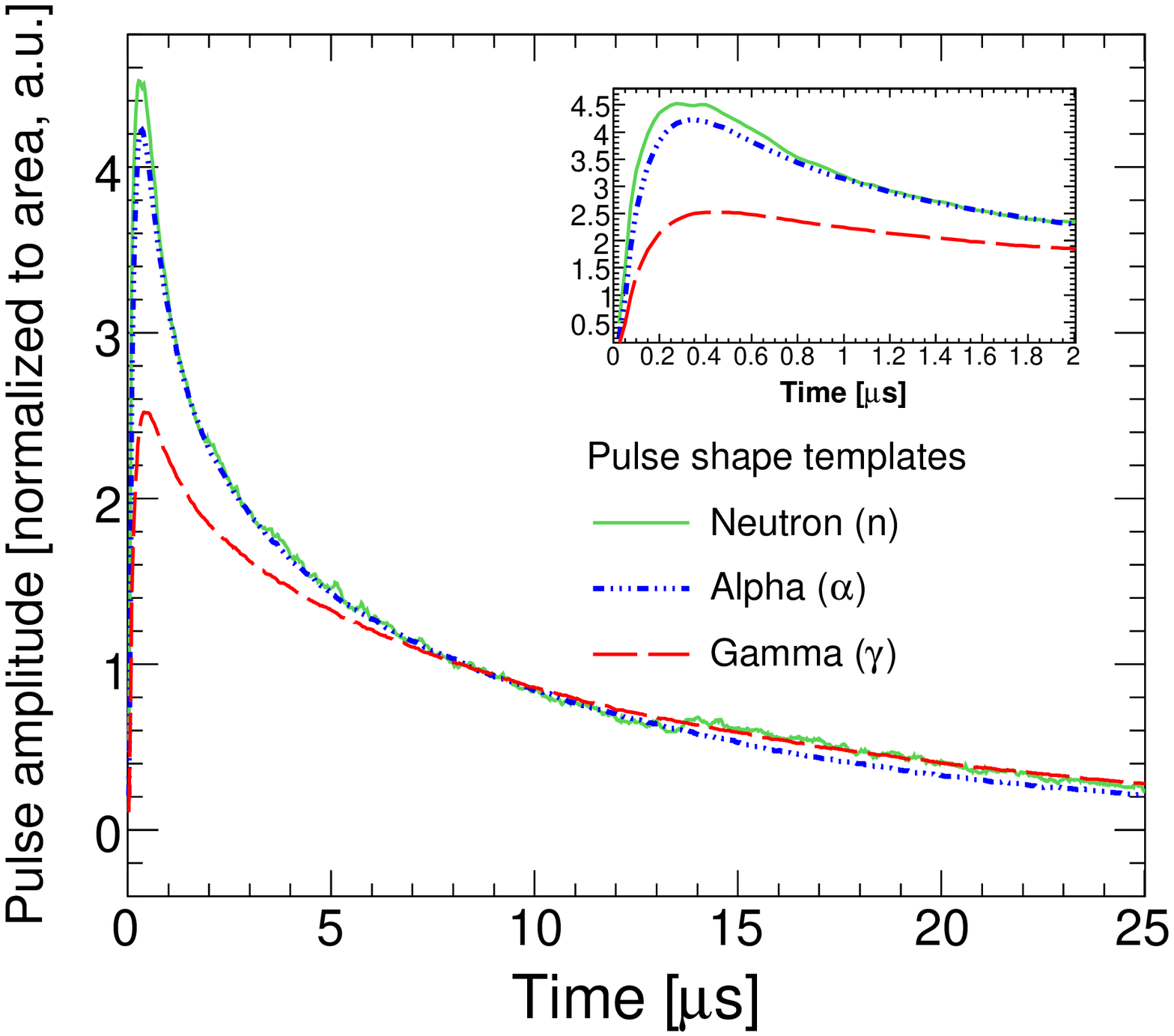}\label{fig:templates}}
\subfigure[]{\includegraphics[width=0.45\textwidth]{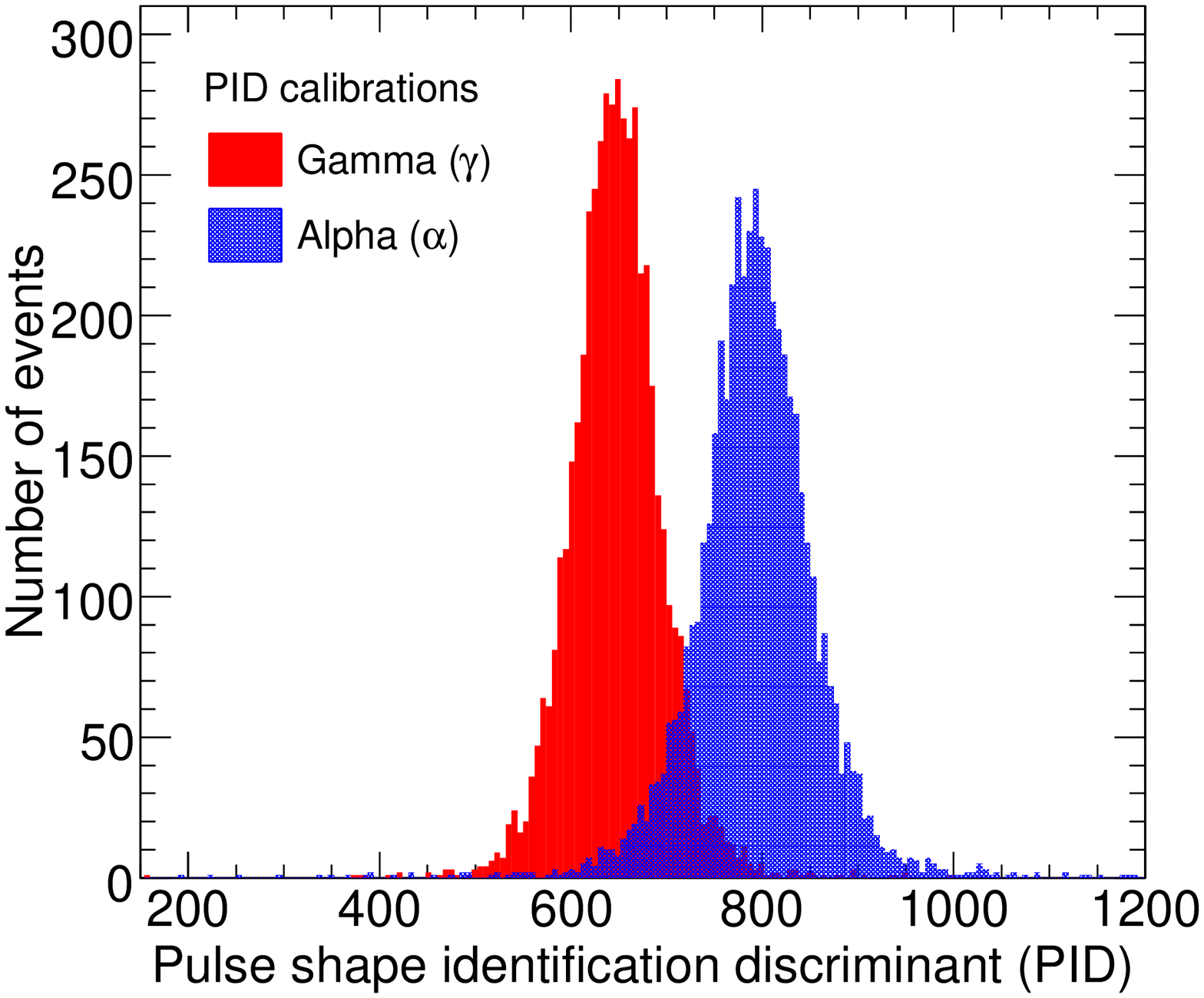}\label{fig:alphaGammaPID}}
\end{center}
\caption{\label{fig:PSDandPID} \subref{fig:templates} Pulse shape ``templates'' for $\alpha$'s, $n$'s, and $\gamma$'s. The emphasis on the fast component of the peak is evident for the $\alpha$ and $n$ recoils. The inset shows the peak region of each pulse shape (0-2 $\mu$s) and highlights the faster rise time observed for the neutron events as compared to $\alpha$ particles. \subref{fig:alphaGammaPID} Pulse shape identification (PID) separation for independent $\alpha$ and $\gamma$ particle calibrations.}
\end{figure*}

We construct the $\gamma$ template with 9322 events using the $^{88}$Y (1.8 MeV) source, the $\alpha$ template with 7781 events from the $^{241}$Am source, and the $n$ template with 450 events from the Am/Be source, each of which are shown in figure~\ref{fig:PSDandPID}. Due to $\gamma$ contamination from the Am/Be source, several tests were performed in order to ensure a high purity of true nuclear recoil events. Layers of polyethylene shielding were added in steps to successively reduce the presence of neutron interactions in the crystal and isolate the $\gamma$ events in the sample. The pulse shape identification (PID) spectrum (shown in figure~\ref{fig:alphaGammaPID} for $\alpha$ and $\gamma$ calibrations and described in more detail below) was then remeasured in order to identify the region populated by the neutron recoils. Events were selected in this range for the measured neutron template. It is important to note here that these templates are obtained from raw data without signal filtering or artificial pulse shaping and represent the CWO response to mechanisms of energy deposited.

The spectral weighting factor is determined using the particle templates described above and is defined as
\begin{equation}
\label{eq:weightingFactor}
\frac{\overline{f}_{\alpha}(t_k)-\overline{f}_{\gamma}(t_k)}
{\overline{f}_{\alpha}(t_k)+\overline{f}_{\gamma}(t_k)}.
\end{equation}
Here, $\overline{f}_{\alpha}$ and $\overline{f}_{\gamma}$ are the pulse amplitudes as a function of time ($t_k$) for the $\alpha$ and $\gamma$ templates, respectively. In order to select events based on PID for the $n$ template, the PID distribution is first calculated using the weighting factor with the $\alpha$ particle template as defined in equation~\ref{eq:weightingFactor}. Once the $n$ template is formed using the selected events, $\overline{f}_{\alpha}$ may be replaced with $\overline{f}_{n}$. We refer to the factor in (\ref{eq:weightingFactor}) as a ``spectral'' weighting factor because it consists of a spectral shape which is to be convoluted with the entire waveform of each event during analysis. The full PID algorithm can be written as
\begin{eqnarray}
\text{PID}=\sum_{k=0}^{k_{f}}\left[\frac{\overline{f}_{\alpha}(t_k)-\overline{f}_{\gamma}(t_k)}
{\overline{f}_{\alpha}(t_k)+\overline{f}_{\gamma}(t_k)}\right]f_{{\text{event}}}(t_k)
\end{eqnarray}
where we have converted this spectral shape in time into a single number, the event PID, by taking the sum over the $k_{f}$ time bins of the event weighted by the weighting factor. Every event is thus assigned a PID value based on this algorithm. To determine the optimum time window ($k_{f}$) over which to integrate the pulse for $\alpha$ and $\gamma$ discrimination, we used the quantity
\begin{eqnarray}
\Delta=\frac{\langle \text{PID} \rangle_{\alpha}-\langle \text{PID}
\rangle_{\gamma}}{\sqrt{\sigma_{\alpha}^2+\sigma_{\gamma}^2}}
\end{eqnarray}
which measures the peak separation of the PID distributions, where $\sigma_{\alpha}$ and $\sigma_{\gamma}$ are the RMS of the PID distributions measured during calibration. This quantity is then maximized for the PID distributions to obtain an optimum time window over which to integrate the pulse of $t_{\rm onset}\leq t_k\leq14{\mu}$s, where $t_{\rm onset}$ is the rising edge of the CWO pulse. This window can be understood qualitatively given the calibration pulse templates in figure~\ref{fig:templates} which converge at approximately 10 $\mu$s.

By using both the above described particle ($\alpha$ and $\gamma$) calibrations, event selection criteria were determined prior to the analysis of the signal data samples and remain consistent throughout the analysis. These criteria are first set using $\gamma$ calibration data to maintain a calibration signal acceptance of greater than 99.7\%, while rejecting electronic noise and square pulses from the livetime pulser. This yields a background $\alpha$ rejection of $\sim$50\%, determined directly from the calibration data (figure~\ref{fig:alphaGammaPID}). Due to the difficulties in obtaining the optimal algorithm for neutron rejection and the iterative procedure used for extracting the neutron pulse shape template, the neutron rejection capability has not been fully quantified for this analysis. The exact fraction of neutron recoils rejected would be exactly characterized with the help of a pure, mono-energetic neutron emitter which was not available at the time of the calorimeter commissioning and operation. 

\subsection{Solar tracking and background data\label{TandB}}

Both background and tracking events are considered for analysis using the same data quality criteria, while solar tracking events have the further requirement that the magnet be sufficiently aligned with the solar core. Corrections are then applied for a small background energy spectrum dependence on the pointing position of the CAST magnet, which is due to differences in natural radioactive background throughout the CAST experimental hall. By dividing the horizontal-vertical plane traversed by the CAST magnet into a set of cells, each of which represents a section of the wall or floor towards which the magnet points at a given time, the position dependence of all detector parameters is directly measured. 

In order to ensure the compatibility of the the final tracking and background spectra, we select background data only from those positions in which tracking data has been recorded and weight those data according to
\begin{equation}
\left(\frac{dN^{\rm BCKG}}{dE}\right)_{\rm eff} = \frac{\sum_{i}(dN^{\rm BCKG}/dE)_it_i}{\sum_it_i}
\end{equation}
where $(dN^{\rm BCKG}/dE)_i$ is the background energy spectrum in the $i^{\textit{th}}$ cell, $t_i$ is the tracking exposure time in the $i^{{\rm th}}$ cell and $(dN^{\rm BCKG}/dE)_{\rm eff}$ is the effective background after position normalization. Following these corrections, the background and tracking (signal) data sets can be reliably compared.

The dataset includes a total of 1257 hours of total exposure time with 60.2 hours of solar alignment and 898 hours of background data. A summary of the statistics for this data set is shown in table~\ref{RunStats}. The effective background data set following the position normalization procedure described above still consists of more than twice the tracking data, thus maintaining good statistics for background subtraction.

\begin{table}
\caption{\label{RunStats} Statistics of the data from CAST Phase~I used in this analysis.}
\begin{indented}
\item[]\begin{tabular}{@{}ll}
\br
Total exposure time & 1257.06 h \\
Solar tracking & 60.256 h \\
Background & 897.835 h \\
BCKG rate after cuts & 1.429 Hz \\
BCKG flux (above 200 keV) & 0.1 \cmSqSec \\
\br
\end{tabular}
\end{indented}
\end{table}

\begin{figure*}
\begin{center}
\subfigure[0.2-3.0~MeV]{\includegraphics[width=0.45\textwidth]{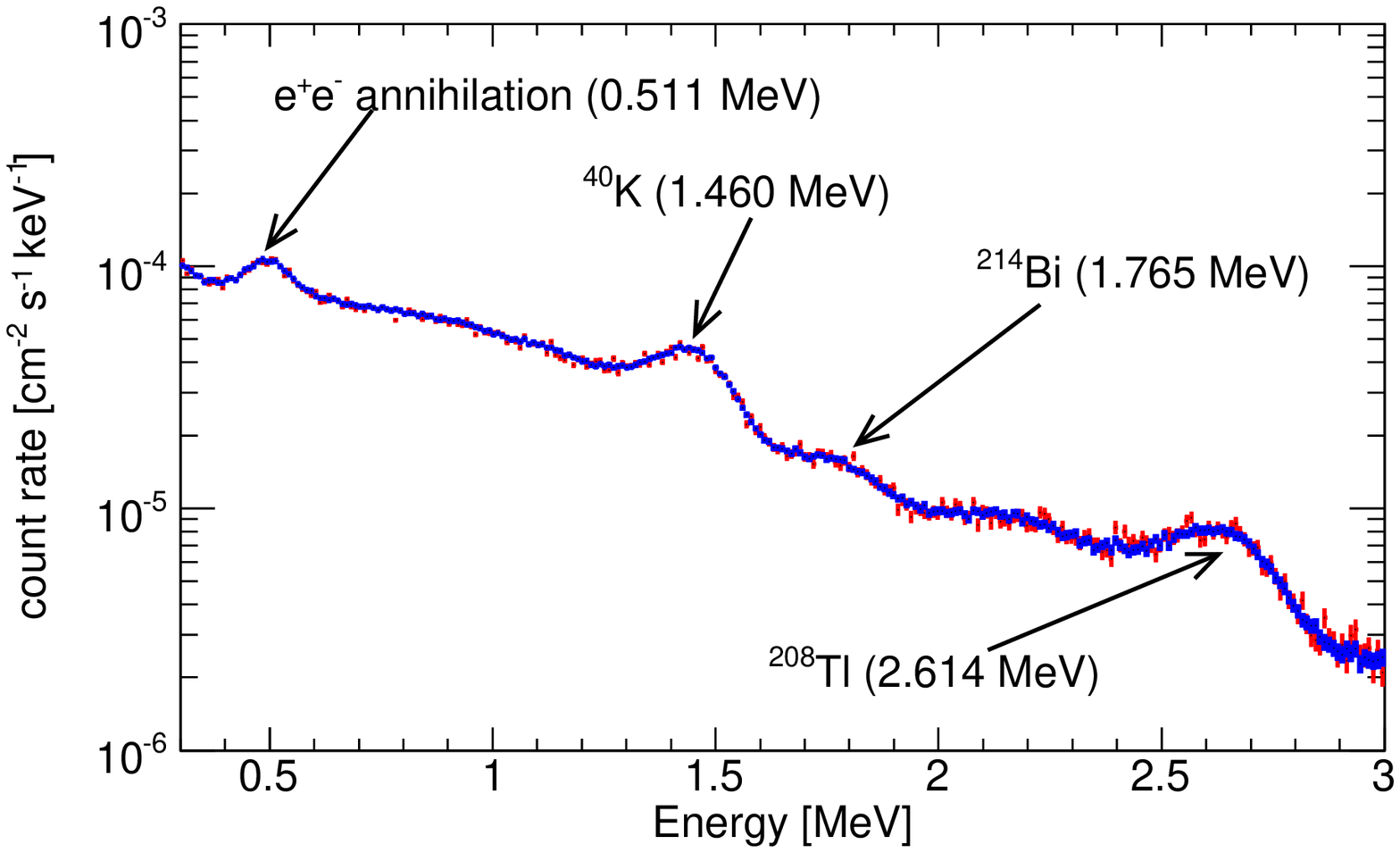}\label{fig:Espectra:low}}
\subfigure[3.0-10~MeV]{\includegraphics[width=0.45\textwidth]{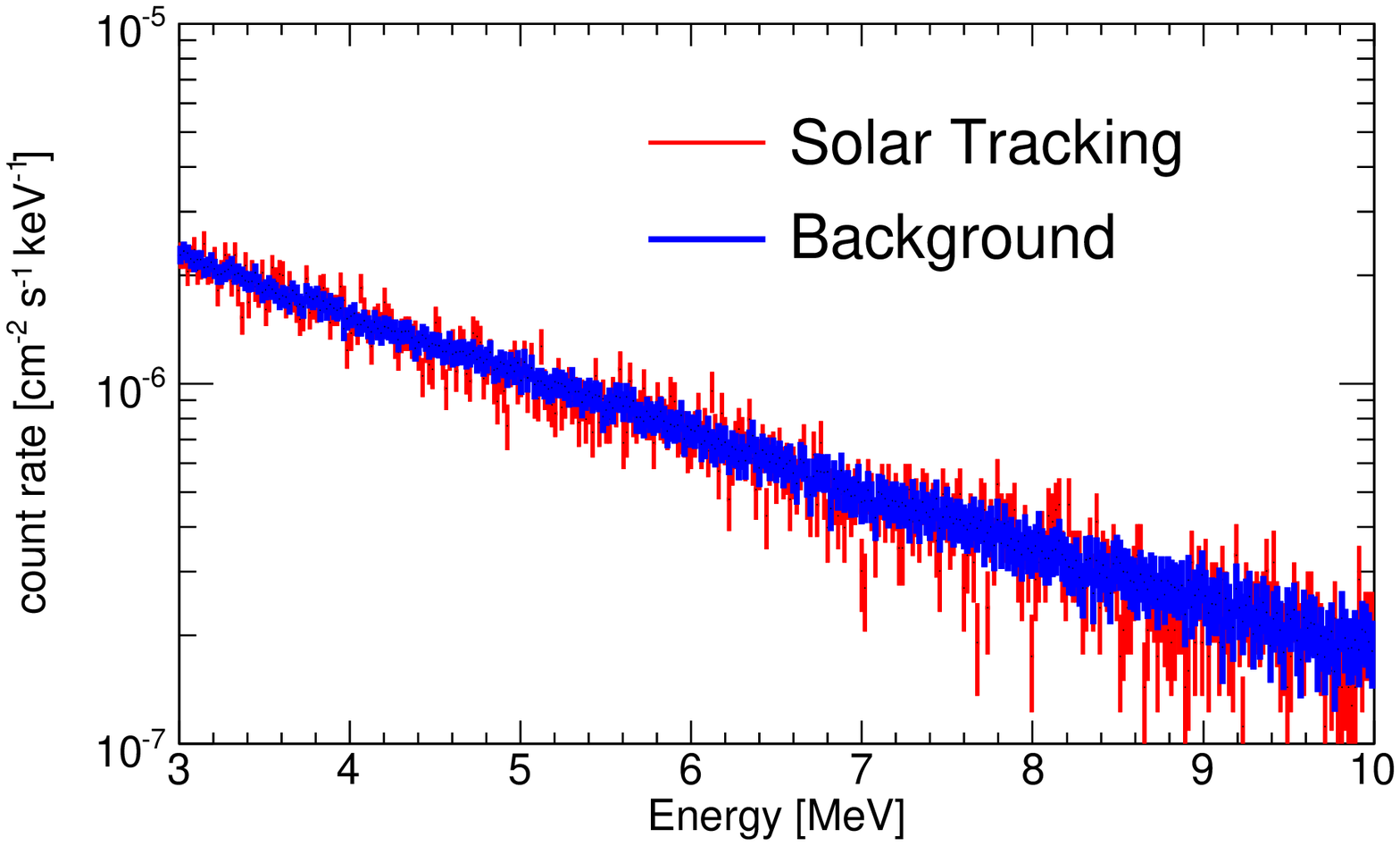}\label{fig:Espectra:mid}}
\subfigure[10-70~MeV]{\includegraphics[width=0.45\textwidth]{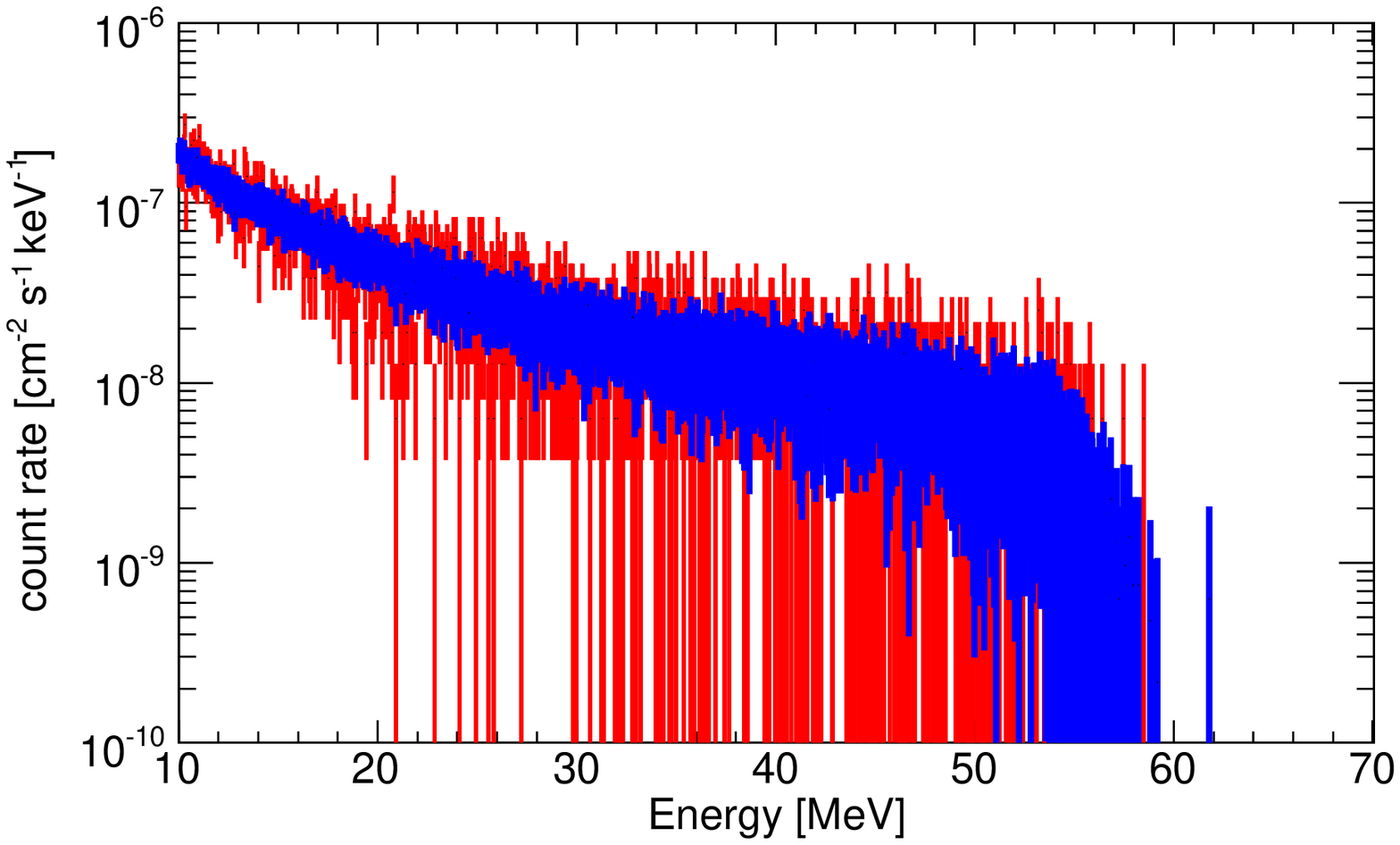}\label{fig:Espectra:hi}}
\end{center}
\caption{\label{Espectra} Tracking and background energy spectra for each energy region in the analysis. The background spectra (narrow band) used have been normalized to the time and position of the tracking data (wider band) and PSD and energy cuts have been applied. Prominent in the low energy region are decays from $^{40}$K activity (1.460~MeV), $^{208}$Tl (2.614~MeV) from the $^{232}$Th decay chain, $^{214}$Bi (1.76~MeV) from the $^{238}$U decay chain, and $e^+e^-$ annihilation $\gamma$'s at 0.511~MeV.}
\end{figure*}

To facilitate the analysis over such the $\gamma$-ray calorimeter large dynamic range in photon energies, the data are divided into three energy regions (0.2-3.0~MeV, 3.0-10~MeV, 10-100~MeV) and binned according to the detector resolution in each region. The energy spectra for both tracking and background in each energy range are shown in figures~\ref{Espectra}. Environmental $\gamma$ radioactivity is very evident in the low energy region. $\gamma$'s from ambient $^{40}$K activity (1.460~MeV) and $^{208}$Tl (2.614~MeV) from the $^{232}$Th decay chain exhibit prominent peaks in the data, along with $e^+e^-$ annihilation $\gamma$'s at 0.511~MeV.

\section{Data analysis and results\label{Analysis}}

\subsection{Expected axion signal \label{Expected}}

Direct background subtraction from the tracking data permits the search for excess events in the residual energy spectrum. The expected signal from axion-photon conversion is a collimated ``beam'' of mono-energetic photons from the magnet bore during solar alignment. This results in Gaussian energy depositions in the CWO crystal for low energy (below 1.022~MeV) photons.

Above 1.022~MeV, an axion conversion photon may pair-produce within the crystal. For each pair-production, there is the possibility that one or both annihilation photons escape. These annihilation escape peaks will lie at 0.511 and 1.022~MeV below the full energy peak and the efficiency for catching these events is characteristic of both the crystal and the energy of the incident axion-conversion photon.

A standard MCNP4b~\cite{mcnp} simulation of this spectrum for a 5.5~MeV photon, convolved with the detector resolution, is used to determine the calorimeter sensitivity to photons at this energy. Photon detection efficiency is nearly 48\% when considering the entire range at 5.5~MeV for this signal (see table~\ref{det}). To validate these data, a laboratory replica of the MICROMEGAS X-ray detector which sits directly in front of the calorimeter in the experiment was constructed and the transmission efficiency through the detector material for photons of various energies was measured and compared to the Monte Carlo predictions. The data and the simulation were found to be in good agreement and the simulation was then used only to determine the peak efficiency and the relative peak heights for the deposition signal. 

The multi-peak signal shape and increased photon detection efficiency improves the sensitivity to excess events above 4.0~MeV. A general search along the entire energy spectrum of the calorimeter would require a full Monte Carlo analysis of the signal shape and its energy dependence. Here, only a 5.5~MeV photon signal has been investigated using this approach, while at all other energies below 10~MeV only a single Gaussian signal (corresponding to a full energy peak) is used. The 5.5~MeV signal is fit as two Gaussians, to a good approximation, which have a fixed peak-height ratio given by the simulation. The search for this signal is described in section~\ref{sect:fuse} and the resulting fit to the data is shown in figure~\ref{residuals}. 

Above 10~MeV, photonuclear dissociation is both energetically possible and very probable, with cross sections near 1 barn for the tungsten and cadmium in the CWO crystal. This large cross section for interaction results in a much different signal shape than for low energy photons and can no longer be approximated by a Gaussian, which is taken into account for the six energies evaluated in the following section. In this energy regime, the total energy deposition efficiency and the signal shape is determined from a standard MCNP4b simulation and depends on the cross-section for photonuclear interactions above 10~MeV. Only 6 points above 10 MeV are evaluated in the model independent scan.

\subsection{\label{sect:scan}Search for anomalous mono-energetic peaks}

We perform a generic search for excess photons across the entire dynamic range of the detector by fitting the known a mono-energetic signal shape to the residual energy spectrum remaining after background subtraction. For the analysis presented here, a Gaussian energy deposition has been used below 10~MeV with the exception of the 5.5~MeV photon signal, as stated in section~\ref{Expected}.

\begin{figure}
\begin{center}
\includegraphics[width=0.65\textwidth]{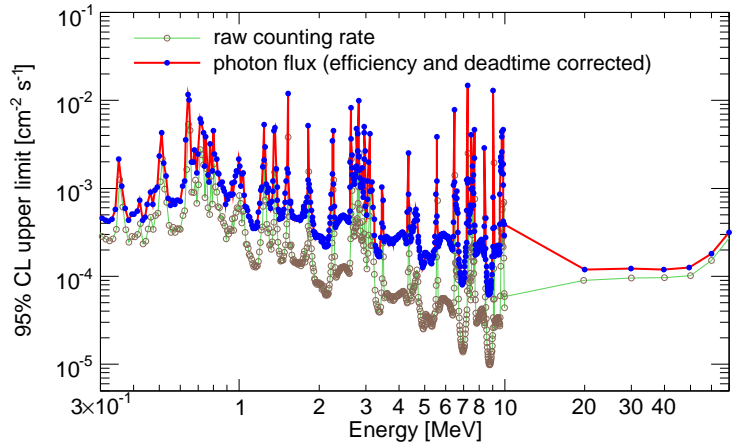}
\caption{\label{AnomalousEvents} 95\% CL limits on the number of counts above background yield limits on the flux of axion-conversion photons incident on the CAST calorimeter after correcting for detection efficiency and livetime. These limits may then be directly translated into the allowed flux of axions incident on the helioscope. Below 10~MeV, a Gaussian energy deposition signal is used. At higher energies, the photonuclear cross section alters the energy deposition signal and the analysis is performed for 20, 30, 40, 50, 60 and 70~MeV.}
\end{center}
\end{figure}

The results of the search and extraction of 95\% CL upper limits on excess photon flux are shown in figure~\ref{AnomalousEvents}. The structure present in the plot is a general consequence of statistical fluctuations in the residual spectrum of the data which lead to large or small 95\% CL bounds on the Gaussian and are physically meaningful as they can point to incomplete subtraction or to slight statistical excesses in the data. Above 10 MeV, 6 points were chosen at which to evaluate the presence of the photonuclear interaction signal (10, 20, 30, 40, 50 and 60~MeV). Although these data are difficult to interpret outside of the context of a particular model, they serve as a benchmark sensitivity for a relatively model independent hadronic axion searches using a helioscope, assuming only that the axion emission is mono-energetic.

\subsection{\label{sect:Li7}Calorimeter sensitivity to \LiSeven}

In order to evaluate the detector sensitivity to axion emission from specific decay channels, several parameters are included and are found in table~\ref{det}.
\begin{table}[!h]
\caption{\label{det}Summary of the calorimeter characteristics and data selection efficiencies used to calculate the expected number of photons incident on the detector for both $^{7}$Li and \HeFuse.}
\begin{indented}
\item[]\begin{tabular}{llcc}
\br
\multicolumn{2}{l}{Production channel} & $^{7}$Li & \HeFuse \\
\mr Peak efficiency & $\Omega_{\mathrm{peak}}$ & \multirow{2}{*}{56.8\%} & \multirow{2}{*}{47.5\%} \\
       Photon transm. efficiency & $\Omega_{\rm transm}$ &  \\
       Energy resolution & $\sigma_{\rm det}$ & 99 keV (21\%) & 327 keV (6\%) \\
       Livetime & $\tau_{\rm det}$ & \multicolumn{2}{c}{93\%} \\
			 Software cuts efficiency & $\Omega_{\rm cuts}$ & \multicolumn{2}{c}{99\%} \\
       Conversion probability & $P_{a\rightarrow\gamma}$ & \multicolumn{2}{c}{(see figure~\ref{convp})} \\
\br
\end{tabular}
\end{indented}
\end{table}
From MCNP Monte Carlo simulations, at the energy of the $^7$Li decay, 478 keV, the efficiency for peak energy deposition after accounting for the photon transmission efficiency through the MICROMEGAS detector is $56.8\%$.
\begin{figure}
\begin{center}
\includegraphics[width=0.65\textwidth]{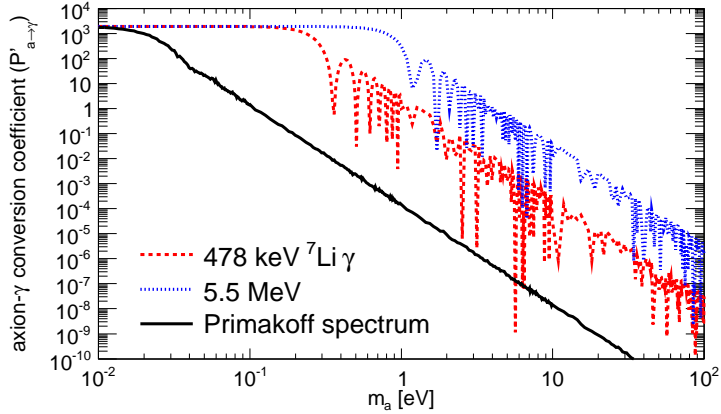}
\caption{\label{convp} Axion-photon conversion coefficient $P'_{\mathrm{a}\rightarrow\gamma}$
versus axion mass. Since axions from nuclear processes have higher energies than Primakoff axions, the sensitivity for axion masses in case of $^{7}$Li decay (478 keV) and p-d fusion (5.5~MeV) is increased for 1 and 2 orders of magnitude, respectively.}
\end{center}
\end{figure}
Including all known detector inefficiency, $\Omega_{\rm total}$, we can estimate the sensitivity to axions from $^7$Li decays using
\begin{eqnarray}
    \Phi_{\rm 478 keV} &=& P_{\mathrm{a}\rightarrow\gamma}(g_{\mathrm{a}\gamma},m_\mathrm{a}, E_\mathrm{a}, B, L)\Phi_{\mathrm{a}}\tau_{\mathrm{det}}\Omega_{\mathrm{total}} \nonumber\\
                          &=& 1.368\times10^{8}P'_{\mathrm{a}\rightarrow\gamma}(m_\mathrm{a}, E_\mathrm{a}, B, L) g^2_{\mathrm{a}\gamma}(g_0+g_3)^2 \hspace{3mm} {\rm cm}^{-2}{\rm s}^{-1}. \label{Li7Lim}
\end{eqnarray}
Here we have factorized the conversion probability $P_{\mathrm{a}\rightarrow\gamma}$ in equation~(\ref{ConvProb}) into the axion-$\gamma$ coupling constant $g_{\mathrm{a}\gamma}$ and the remaining numerical term, $P'_{\mathrm{a}\rightarrow\gamma}(m_{\mathrm{a}}, E_{\mathrm{a}}, B, L)$, which only depends on the magnet parameters and the axion energy and mass. $P'_{\mathrm{a}\rightarrow\gamma}(m_{\mathrm{a}}, E_{\mathrm{a}}, B, L)$ versus $m_{\mathrm{a}}$ is shown in figure~\ref{convp}.

\begin{figure*}
\begin{center}
\subfigure[]{\includegraphics[width=0.45\textwidth]{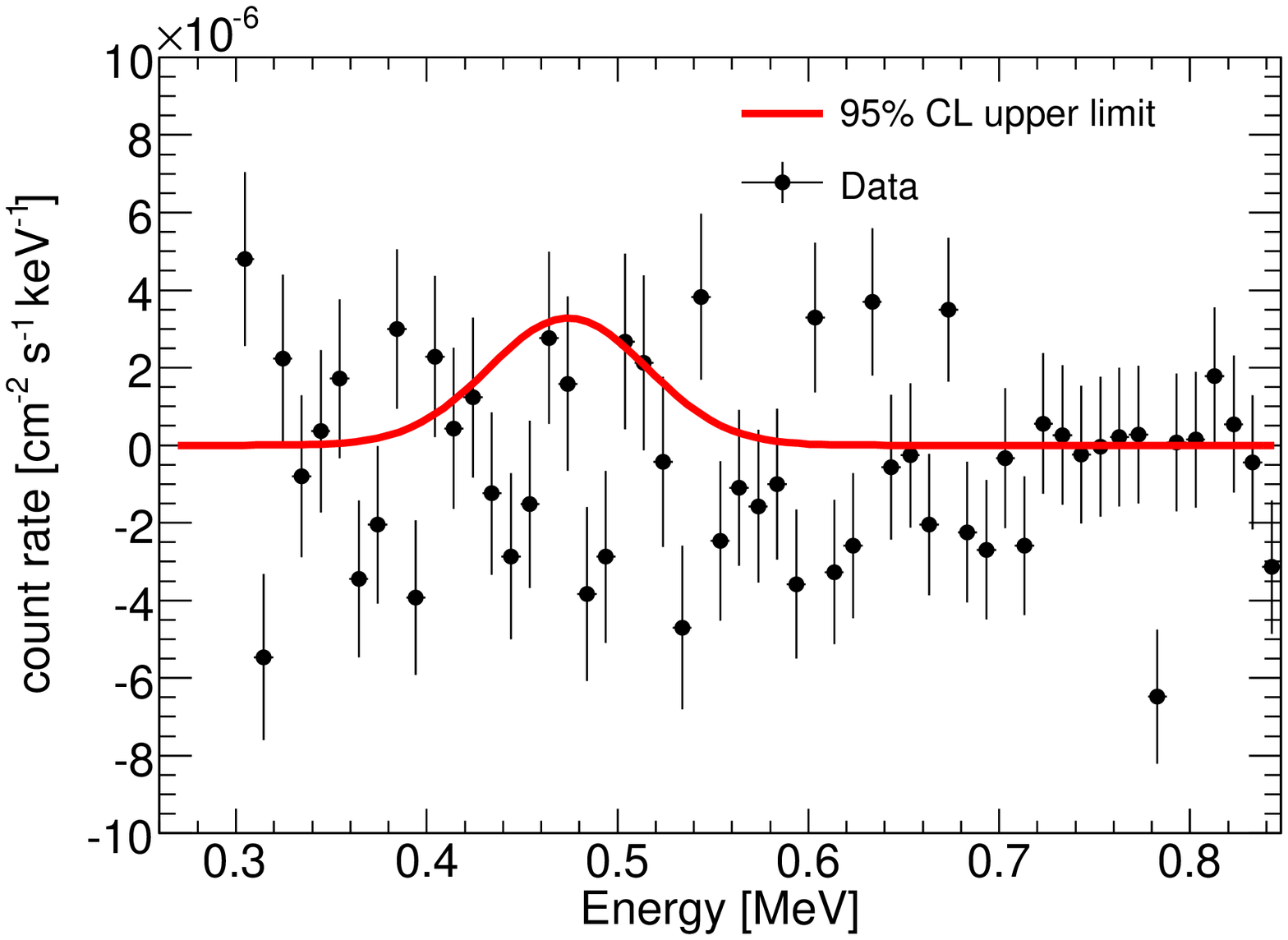}\label{fig:residuals:Lithium}}
\subfigure[]{\includegraphics[width=0.45\textwidth]{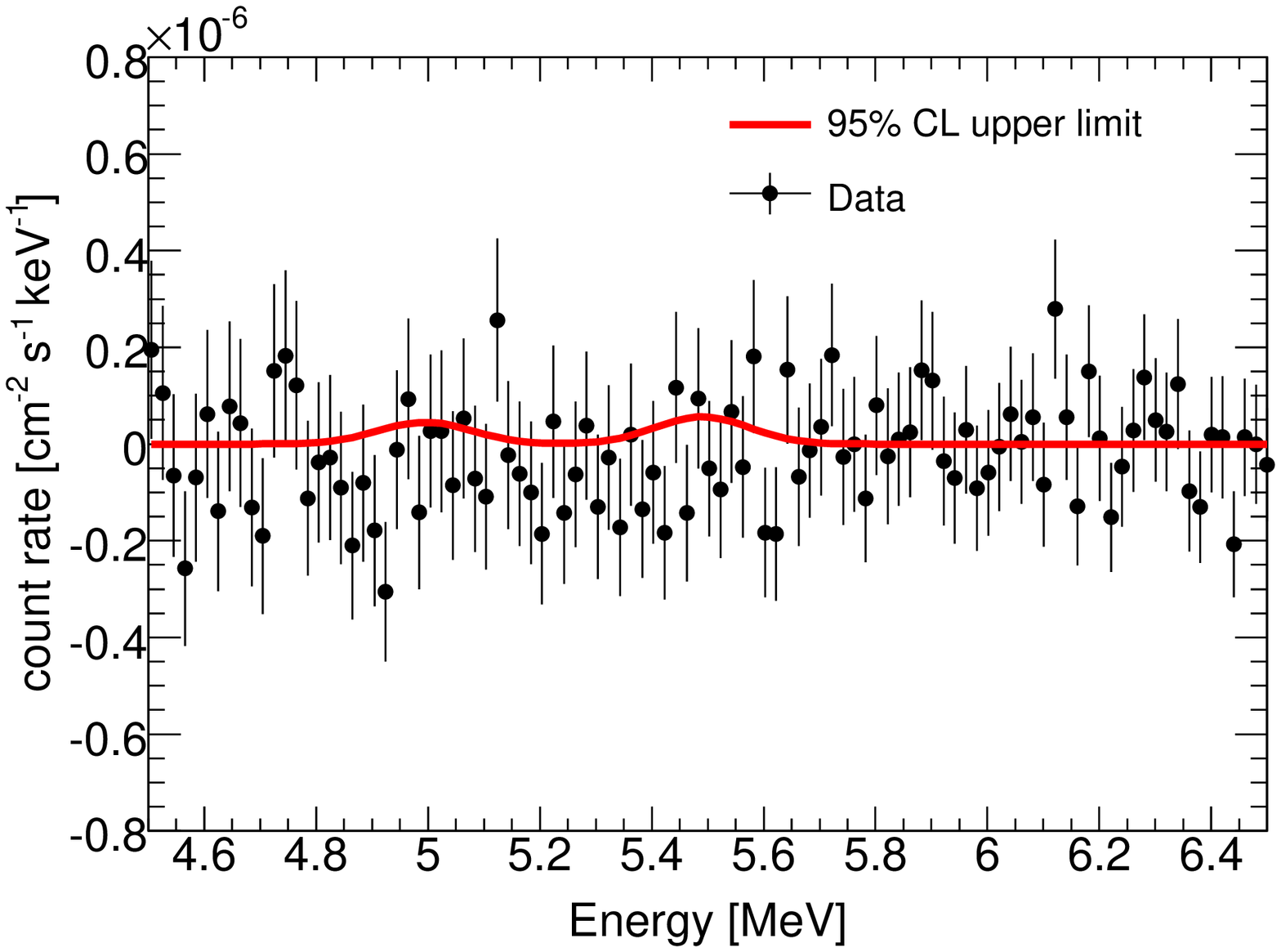}\label{fig:residuals:Helium}}
\end{center}
\caption{\label{residuals} Fits to the known signal shape after background subtraction. \subref{fig:residuals:Lithium} A single Gaussian fit for \LiSeven\ results in a 95\% CL upper limit for a 478 keV axion-conversion photon of $\Phi_{\rm 478 keV}(95\%{\rm CL})\le3.4\times10^{-4}$ \cmSqSec. \subref{fig:residuals:Helium} A double Gaussian fit to the expected signal from \HeFuse\ yields the 95\% CL upper limit for a 5.5~MeV axion-conversion photon of $\Phi_{\rm5.5 MeV}(95\%{\rm CL})\le2.14\times10^{-5}$ \cmSqSec.}
\end{figure*}

\begin{figure}
\begin{center}
\includegraphics[width=0.75\textwidth]{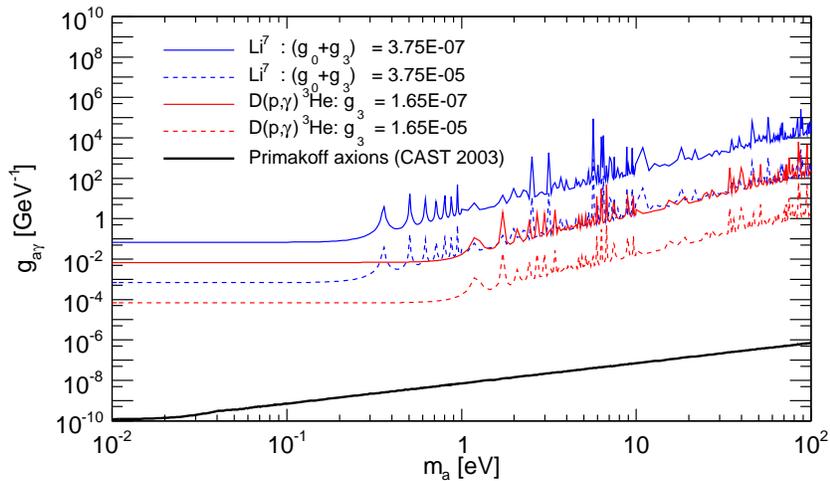}
\caption{\label{limits} The limits obtained on the axion-photon coupling versus axion rest mass for 478 keV axions from $^7$Li decay and 5.5~MeV axions from proton-deuteron fusion for two values of the nuclear couplings. This parasitic axion search has not found any evidence for new pseudoscalar particles coupling to nucleons.}
\end{center}
\end{figure}

To extract the signal, the background spectrum was subtracted from the spectrum obtained during solar alignment, and the resulting residual is shown in figure~\ref{fig:residuals:Lithium}. Since no evidence of an axion signal was observed only an upper limit could be determined. A fit to the expected Gaussian signal shape yields a 95\% CL upper limit on excess photon events at 478 keV of $\Phi_{\rm 478 keV}(95\%{\rm CL})\le3.4\times10^{-4}$ \cmSqSec. 
By solving for $g_{\mathrm{a}\gamma}$ in (\ref{Li7Lim}) we can write
\begin{eqnarray}
    g_{\mathrm{a}\gamma}&\le&\frac{1}{g_0+g_3}\sqrt{\frac{\Phi_{\rm 478 keV}(95\%{\rm CL})}{1.368\times10^{8}}\frac{1}{P'_{\mathrm{a}\rightarrow\gamma}}}. \label{limiteq}
\end{eqnarray}
The limits as a function of $m_{\mathrm{a}}$ are shown in figure~\ref{limits} for two different values of the nuclear coupling constant ($g_0+g_3$). These values are chosen and evaluated using equation~(\ref{g3}) as representative of the range of couplings for a Peccei-Quinn scale of $f_{\rm a}=10^6-10^8$~GeV.

\subsection{\label{sect:fuse}Calorimeter sensitivity to \HeFuse}

Using again the information from table~\ref{det}, we can evaluate the limiting expression for $g_{\mathrm{a}\gamma}$ in the \HeFuse\ channel using
\begin{eqnarray}
    \Phi_{\rm 5.5 MeV} &=& 0.8878\times10^{10}P'_{\rm{a}\rightarrow\gamma}(m_{\rm{a}}, E_{\rm{a}}, B, L) g^2_{\rm{a}\gamma}g_3^2 \hspace{3mm}{\rm cm}^{-2}{\rm s}^{-1} . \label{DfuseLim}
\end{eqnarray}
The resulting spectrum after background subtraction is shown in figure~\ref{fig:residuals:Helium}.
For this reaction, the expected signal differs from that of $^7$Li since a 5.5~MeV $\gamma$-ray can pair-produce within the calorimeter. This escape peak structure has been taken into account, including the fixed peak-height ratio, resulting in a 95\% CL limit on excess photons of $\Phi_{\rm 5.5 MeV}(95\%{\rm CL})\le2.14\times10^{-5}$ events \cmSqSec. Using equation~(\ref{DfuseLim}) we have
\begin{equation}
    g_{\rm{a}\gamma}\le\frac{1}{g_3}\sqrt{\frac{\Phi_{\rm 5.5 MeV}(95\%{\rm CL})}{0.8878\times10^{10}}\frac{1}{P'_{\rm{a}\rightarrow\gamma}}}. \label{DFuseCoupLim}
\end{equation}

We again use the two different values of the nuclear coupling constant ($g_3$), evaluated using equation~(\ref{g3}) and a Peccei-Quinn axion scale of $f_{\rm a}=10^6-10^8$ GeV. By plotting the axion-photon coupling constant versus axion mass, we see that the limits are weaker than those obtained by the CAST X-ray detectors in 2003. 

\section{Conclusions}


The CAST photon calorimeter provides a search for high-energy axion-photon conversions during periods of solar alignment. This is the first such search for high-energy pseudoscalar bosons with couplings to nucleons conducted using a helioscope approach and provides an important cross-check for other searches focused on nuclear decay, such as~\cite{krc01,bellini08}. Furthermore, as discussed in~\cite{raf82}, the search for pseudo-scalar emission from proton-deuteron fusion (\HeFuse) is potentially sensitive to a more general class of new particles than only Primakoff or hadronic axions due to the presence of both M1 and E1 transitions and can couple particles of various spin-parity. 

In making use of the CAST magnet for an axion search strategy not initially foreseen, the achievable sensitivity is severely limited and a dedicated high-energy axion search performed underground and without shielding limitations would be able to reach background levels many orders of magnitude lower. Such levels are necessary to reach the very small axion flux expected from the two axion emission channels considered in this search. CAST remains a unique instrument with unprecedented sensitivity allowing for new searches for anomalous solar emissions in the form of new axion-like particles with coupling to photons.

\section*{Acknowledgements}
We would like to thank CERN for making this experiment possible. We acknowledge support from NSERC (Canada), MSES (Croatia) under the grant number 098-0982887-2872, CEA (France), BMBF (Germany) under the grant numbers 05 CC2EEA/9 and 05 CC1RD1/0 and DFG (Germany) under grant number HO 1400/7-1, the Virtuelles Institut f\"ur Dunkle Materie und Neutrinos -- VIDMAN (Germany), GSRT (Greece), RFFR (Russia), the Spanish Ministry of Science and Innovation (MICINN) under grants FPA2004-00973 and FPA2007-62833, NSF (USA) under Award number 0239812, US Department of Energy, NASA under the grant number NAG5-10842 and the helpful discussions within the network on direct dark matter detection of the ILIAS integrating activity (Contract number: RII3-CT-2003-506222).

\section*{References}


\end{document}